\documentclass[10pt, conference, letterpaper]{IEEEtran}
\usepackage{graphicx}
\usepackage{xcolor}
\usepackage{amsmath}
\usepackage{tikz}
\usepackage{pgf-umlsd}
\usetikzlibrary{shapes,arrows,shadows} % for pgf-umlsd
\usepackage{amsthm}
\usepackage{bm}
\usepackage{tikz}
\usepackage{forest}
\usepackage{flushend}
\usepackage[font=small,labelfont=bf]{caption}
\usepackage{subcaption}
\usepackage{amssymb}
\usepackage{subcaption}
\usepackage{algorithm}
\usepackage[noend]{algpseudocode}
\usepackage{pgf}
\usepackage[left=0.65in,right=0.65in,top=0.75in,bottom=1in]{geometry}
\usetikzlibrary{arrows,automata,shadows}
\usepackage[latin1]{inputenc}
\usetikzlibrary{decorations.pathreplacing}

\DeclareMathOperator*{\expectation}{E}
\newcommand{\Exp}[1]{\expectation \left[ #1\right]}
\graphicspath{{./images/}}

 \theoremstyle{definition}
\definecolor{TUMBlau}{RGB}{0,101,189} % Pantone 300
\definecolor{TUMBlauDunkel}{RGB}{0,82,147} % Pantone 301
\definecolor{TUMBlauHell}{RGB}{152,198,234} % Pantone 283
\definecolor{TUMBlauMittel}{RGB}{100,160,200} % Pantone 542

\definecolor{TUMElfenbein}{RGB}{218,215,203} % Pantone 7527 -Elfenbein
\definecolor{TUMGruen}{RGB}{162,173,0} % Pantone 383 - Gr?n
\definecolor{TUMOrange}{RGB}{227,114,34} % Pantone 158 - Orange
\definecolor{TUMGrau}{gray}{0.6} % Grau 60%

\usepackage[absolute,showboxes]{textpos}

\TPMargin{5pt}

\definecolor{gl}{rgb}{0.0,0.5,0.8}
\definecolor{fc}{rgb}{0.8,0.5,0}
\definecolor{al}{rgb}{1,0.3,0.3}

\begin{document}
	%\copyrightstatement
%
% paper title
% Titles are generally capitalized except for words such as a, an, and, as,
% at, but, by, for, in, nor, of, on, or, the, to and up, which are usually
% not capitalized unless they are the first or last word of the title.
% Linebreaks \\ can be used within to get better formatting as desired.
% Do not put math or special symbols in the title.

\title{Hard Latency-Constraints for \\ High-Throughput Random Access: SICQTA}

\author{\IEEEauthorblockN{H. Murat G\"ursu, Fuqi Guan,  Wolfgang Kellerer}
\IEEEauthorblockA{Chair of Communication Networks, Technical University of Munich, Munich, Germany\\ %\IEEEauthorrefmark{1}
%\IEEEauthorrefmark{2}Department of Electronic Systems, Aalborg University, Aalborg, Denmark \\
E-mail:\{murat.guersu, fuqi.guan, wolfgang.kellerer\}@tum.de}}

% make the title area
\maketitle

% For peer review papers, you can put extra information on the cover
% page as needed:
% \ifCLASSOPTIONpeerreview
% \begin{center} \bfseries EDICS Category: 3-BBND \end{center}
% \fi
%
% For peerreview papers, this IEEEtran command inserts a page break and
% creates the second title. It will be ignored for other modes.
\IEEEpeerreviewmaketitle

\begin{abstract}
Enabling closed control loops via wireless communication has attracted a lot of interest recently and is investigated under the name cyber-physical systems. Under cyber-physical systems one challenging scenario is multiple loops sharing a wireless medium, and the age of the control information has to be minimized without sacrificing reliability to guarantee the control stability. The number of transmitting devices depends on the control parameters thus, it is stochastic. Wireless uplink resource allocation given low latency constraints for unknown number of devices is a hard problem. For this problem, random access is the most prominent way to minimize latency, but reliability is sacrificed. However, as reliability is also critical for such applications, improved random access algorithms with hard latency guarantees are needed. Currently available random access algorithms with hard latency guarantees have low throughput and some of them are limited to low number of active devices. In this work, we provide a high-throughput random access algorithm with hard latency-constraints (SICQTA) that scales to any number of active devices. This algorithm, making use of feedback, has a varying throughput between $0.69$ and $1$ depending on the number of devices, which is unprecedented in the state of the art up to our best knowledge.
\end{abstract}

\section{Introduction \& Background}
\label{sec:intro}

%Discussion on latency-reliability requirements.
%\mg{Possibly write another introduction focusing on control systems.}

%\mg{I believe CPS can be a really good use-case for the capability of the algorithm. The algorithm shines if the total number of devices $N$ is relatively small $<100$ and active number of devices $M$ varies. Shall I start from 5G to go to CPS. Or directly motivate through CPS? }
%Upcoming exciting use-cases such as autonomous driving, industrial automation and emergence of robotics all around us have been pushing the communication research for more stringent latency-reliability constraints \cite{keylist}. These requirements cannot be fulfilled with scheduled access anymore due to overhead of signaling for uplink communication. Grant-free is proposed as an alternative due to flexibility of allocating resource to devices \cite{keylist}. Even though the signaling overhead is overcome, the reliability of communication is at stake and recent work have emphasized on delivering reliable grant-free schemes \cite{keylist}.

One typical problem with latency-reliability constraints is uplink resource allocation for cyber physical systems \cite{tabuada2007event}. In this problem, multiple control loops share the wireless medium. Each loop is composed of a controller, actuator and a sensor. The controller is located at a central entity while actuator and sensor are both located at the device. The closed control loops outputs actuation decisions in the controller from the input of the sensing information. The devices transmit the sensing information through uplink communication and get actuation decision as downlink communication. 

\par The downlink is broadcast to all actuators without the need of coordination. However, depending on the state of the control loop, only some of the sensors transmit state information through uplink communication. As the transmission depend on the state of the control, the number of devices transmitting at a certain time is unknown. Thus, we have $M$ active sensors at a certain time out of $N$ total sensors which have to be allocated resources to optimize the control performance. This problem is previously investigated with LTE scheduling consisting of a scenario with multiple inverted pendulums in \cite{vilgelm2017control}. However, the solution assumes the information of device activity to overcome the over-dimensioning of scheduling. This information is not available in reality and the inefficiency to obtain this information has actually called for a new design of LTE uplink resource allocation mechanism called as grant-free \cite{RP_181477}, reusing the state of the art in random access area.

Grant-free focuses on a scenario where devices transmit a single packet or multiple replicas to achieve the latency-reliability constraints. This requires over-dimensioning of resources to fulfill tight reliability constraints as it lacks the information that is the number of active devices \cite{gursu2018multiplicity}. As a solution to over-dimensioning, successive interference cancellation (SIC) is integrated to the random access schemes.

%\mg{paragraph on SIC schemes and how they increase throughput but each one has its own drawbacks.}
SIC enables recovery of overlapping packets through signal processing. This has increased the throughput of random access algorithms from $0.5$ packets per slot up to $1$ packet per slot with asymptotic number of devices, reaching the efficiency of scheduling based solutions. The trade-off is the decoding complexity. Through edge-cloud processing and distributed computing, complexity is expected to be dealt with for radio access algorithms \cite{el2018edge}.

Successive interference cancellation is initially explored for tree algorithms in \cite{yu2005sicta}. Through that work the throughput for tree algorithms is increased to $0.69$ from $0.35$. In \cite{yu2005sicta} the clean packet for cancellation is guaranteed with feedback, forcing devices to split from each other. However, too much structure is inefficient and in \cite{liva2011graph} it is shown that the same structure can be built through random decisions. The random decisions are shaped with a degree distribution tailored to the number of devices. It is shown that the algorithm reaches a throughput of $1$ in the asymptotic region when $M$ goes to infinity. %This work is extended from transmission of packets to transmission of codewords in \cite{paolini2015coded}. 

%The formation of a tree structures is necessary for the successive nature of the interference cancellation. Through his expertise on LDPC codes, Gianluigi Liva have seen that the trees needed for SIC is not necessarily formed through feedback but can be formed through random decisions in \cite{liva2011graph}.

Another work \cite{stefanovic2012frameless} adapts that work to a frameless structure where the degree distribution is replaced with setting a Binomial probability to transmit at each slot. Compared to framed structure the results show that, \cite{stefanovic2012frameless} has a better performance in the non-asymptotic region. However, neither of these algorithms can provide a hard guarantee on the latency. Also both of them are susceptible to varying number of active devices. The hard guarantees can be provided via setting the decisions uniquely for each device. 

\par This problem is initially investigated by Massey under the name "protocol sequences" for de-synchronized devices in \cite{massey1985collision}. These algorithms are too pessimistic to be applied to tight latency constraints as the offset between devices is the main issue there and it is not the main problem anymore thanks to improvement in hardware design. Recently, the unique decisions for each device for hard guarantees is investigated in \cite{Boyd2018} under the name "access codes", where each device transmits packets with respect to a unique code. The design of these codes is of combinatorial complexity. The results are limited, as we detail on later parts of the paper. Moreover, the use of feedback is neglected in this work.
%Some codes for $2$ and $3$ active devices are shared and other values are left out due to complexity of the problem. 
\par Uniqueness of the access decisions can be guaranteed through feedback to overcome the complexity of proposed protocol. Using addresses for such limitation is initially proposed by \cite{capetanakis1979tree} and adapted for RFID tags with Query Tree Algorithms in \cite{choi2007query}. However, this algorithm lacks behind in throughput compared to SIC-capable algorithms. The idea to use Interference Cancellation for Query Tree Algorithms is introduced in \cite{kumar2011interference}. However, the explanation of the algorithm in \cite{kumar2011interference} is unclear. The throughput they have shown is capped to $0.69$ which have already been shown by \cite{yu2005sicta} for TA with SIC capabilities. Hard guarantees for performance are not investigated and the difference to \cite{yu2005sicta} is unclear. 
  
In our work we propose a novel Successive Interference Cancellation for Query Tree Algorithm, SICQTA. We provide analytical hard upper and lower bounds to latency and compare it with simulations to show the validity. It is shown that the algorithm easily extends to any number of active devices unlike \textit{access codes}, and it provides a higher throughput compared to previous SIC based works. On top of that, hard latency guarantees make it a suitable candidate as a solution of the uplink resource allocation problem with unknown number of active devices. % On top of that we provide a tighter upper bound for latency of QTA. 
%\mg{this probably needs to be re-written.}

 Our paper is organized as follows: In Sec.~\ref{sec:define} we explain the scenario and provide the problem formulation for reliable access with latency constraints. In Sec.~\ref{sec:det_tree} we introduce shortly the Query Tree Algorithm and Successive Interference Cancellation Query Tree Algorithm. In Sec.~\ref{sec:evalu} the latency bounds are given and we compare our solution to the \textit{access codes} while comparing the bounds with simulations. Further discussions are given in Sec.~\ref{sec:disc}. Finally, the paper is concluded with possible extensions of future work in Sec.~\ref{sec:conc}. 

\vspace{-0.1cm}
\section{Scenario \& Problem}
\label{sec:define}
\vspace{-0.1cm}
\par %The definition of a resource converts the wireless communication problem to a resource allocation problem between a set of transmitters and receivers.
 We consider a star topology where the central entity is called the gateway and leaf entities of the star are called devices. We consider an uplink scenario where only devices transmit a packet to the gateway. There are $N$ devices attached to the gateway. Considered resources in the system are slots of a single channel with a TDM scheme. 
 
 \par Two different channel models are considered with and without SIC. First one is a collision channel model where perfect reception is assumed. If there is no contention, there is no loss of packets \cite{ghez1988stability}. Second one is for SIC scenario, we assume perfect cancellation is possible if clean packets are received. These assumptions are common in MAC layer research to focus on a layer 2 based solution. Impact of more practical channel models are discussed in Sec.~\ref{sec:disc}. Each device is synchronized perfectly to the slots defined by the TDM structure. The devices are randomly and sporadically activated and the number of active devices at any slot is $M$, such that $M\leq N$. The devices have a homogeneous radio latency constraint $\text{L}$ and reliability constraint $\text{R}$. We investigate the multiple access problem of maximizing throughput that we abstract as maximizing number of successfully used slots.\footnote{For simplicity we assume that the constraint can be expressed in terms of slots. The reliability constraint here is the radio layer reliability, that can be input to the end to end reliability model.}
 \par We define a frame structure consisting of $d$ subsequent slots. We investigate the problem of designing codes $\mathbf{c}$ that represents the binary access decision of a device. The code $\mathbf{c}$ is of size $d$, i.e., $\mathbf{c} = \{c_1, c_2, \cdots , c_d\}$ where $c_i \in \{0,1\}\, \forall\, i \in\{1,\dots,d\}$. The device that has the code $c_i=1$ will transmit its packet at slot $i$. %This problem is initially introduced by Massey in \cite{massey1985collision} under the name \textit{protocol sequences}.

The codebook $\mathbf{C}$ is a collection of all codes and is a matrix with $d$ columns and $N$ rows where each row represents a unique code for each device. An example is as follows:

\[
\mathbf{C}=
\begin{bmatrix}
0 & 0 & 1 & 0 \\
0 & 1 & 0 & 0 \\
1 & 0 & 0 & 0
\end{bmatrix}
\]
%If all the devices are active at the same time 
with $N=3$ devices and a frame size $d=4$. Each device is sporadically active and the activity of all devices is represented with a vector $\mathbf{n}$ with $N$ elements, i.e., $n_j \,\in \{0,1\} $ where $n_j=1$ represents that the device $j$ is active. We assume that codebook $\mathbf{C}$ is ordered such that code of device $j$ is in the $j^\text{th}$ row of $\mathbf{C}$. This assumption allows us to define a frame outcome $\mathbf{f}$ as in,
\begin{equation}
\mathbf{n}\cdot \mathbf{C}= \mathbf{f}. 
\end{equation}
The frame outcome  $\mathbf{f}$ represents the number of packets at each slot of the frame. However, receiver is unaware of this information such that $\mathbf{f}$ should be converted to MAC layer success outcome $\mathbf{s}$. An example for collision channel would be,
\vspace{-0.1cm}
%$K$ multi-packet reception capability (K-MPR) \cite{keylist}. 
\begin{equation}
f_i  
\begin{cases}
= 1 & s_i = 1 \\
\text{o.w.} &  s_i = 0.
\end{cases}
\end{equation}
\vspace{-0.1cm}
%where the success outcome $\mathbf{s}$ is the converted version of the frame outcome. The outcome is successes on resource 1 and 2.  
%\[
%\begin{bmatrix}
%0  & 1 & 1 
%\end{bmatrix}
%\begin{bmatrix}
%0 & 1 & 1 & 0 \\
%0 & 1 & 0 & 0 \\
%1 & 0 & 0 & 0
%\end{bmatrix} 
%= 
%\begin{bmatrix}
%1  & 1 & 0 & 0 
%\end{bmatrix}.
%\]
%The frame outcome $\mathbf{f}$ can be represent with rowwise summation of the matrix $\mathbf{C}$. The active devices can be represented with an activation vector $\mathbf{a}$ where each active device is represented with a one and vice versa i.e., $\mathbf{a} = [ 0 \, 1\, 1 ]$ and $a_j \in \{0,1\}\, \forall\, j\, \in \{1,2,...,N\}$. When the first device is inactive and the others are active, the left hand side multiplication of the resource allocation matrix with the activity vector results in the frame outcome $\mathbf{f}$,
%\[
%\mathbf{a}\cdot \mathbf{C}= \mathbf{f} = 
%\begin{bmatrix}
%0  & 1 & 1 
%\end{bmatrix}
%\begin{bmatrix}
%0 & 0 & 1 & 0 \\
%0 & 1 & 0 & 0 \\
%1 & 0 & 0 & 0
%\end{bmatrix} 
%= 
%\begin{bmatrix}
%1  & 1 & 0 & 0 
%\end{bmatrix}
%\]
% The frame outcome for $i^{\text{th}}$ resource  $f_i \in \mathbf{f}$ considering the  collision resource model constraint can be re-written as
Using the previous definitions we can define an optimization problem for codebook design.

%\subsection{Problem Definition} %With $M$ active devices and 
 Given frame size $d$ and number of devices $N$, maximize the total success per frame through the codebook design $\mathbf{C}$:%For the deterministic access, the resource allocation matrix $\mathbf{C}$ has to obey the following rule: Given any $\mathbf{a}$ such that $\| \mathbf{a} \|^2 = M$, the $\mathbf{a}\cdot \mathbf{C}= \mathbf{f}$ that is re-written as $\mathbf{s}$ has to obey the following $\| \mathbf{s} \|^2 = M$. 
%If $M = N$ all the devices are active all the frames and we should allocate one resource for each device. We call this scenario scheduling    %In certain scenarios the device has a binary payload such that the transmission of the unique idendit
%\mg{not applicable to K-MPR think on it.}
\begin{align}
	\arg \max_\mathbf{C} & \|  \mathbf{s} \|^2,%\,\,\,\,\, 	\text{subject to: }  = M, \,\,\, #
	\\\text{s.t. }  \mathbf{n} &\in \mathcal{N},
	\\  d&\leq \text{L},
	\\  \Exp{\|  \mathbf{s} \|^2 }&\geq\text{R} \Exp{\|  \mathbf{n} \|^2}. \label{eq:last}	
	% =\mathbf{f}   \,\,\, \text{, }    s_i = \delta_{f_i,1} \text{ and } \| \mathbf{a} \|^2 = M
\end{align}
where $\mathcal{N}$ is the set of all possible activation combinations of $N$ devices, L and R are the latency and reliability constraint respectively. The $\|.\|^2$ operation is the autocorrelation operation that also gives the summation of binary vectors. This is naturally a combinatorial problem and hard to solve, as $\mathbf{n}$ can take any value. We can write  $\|  \mathbf{n} \|^2 = M$, where $M$ is the number of simultaneously active devices per frame. 
\par The problem definition is shared here for formalism. In the following part of the paper we show that SICQTA solves this problem with a distributed algorithm that is guided via a central feedback. Optimality of the algorithm is not proven is an open issue for future work.

\section{Algorithms with Feedback}
\label{sec:det_tree}
\begin{figure*}[!htb]
	\centering
	\begin{subfigure}[t]{0.49\textwidth}
			\centering
		\begin{tabular}{|c|c|c|c|c|c|c|c|c|c|c|}
			\hline
			\scriptsize{A,B,C,D} & \scriptsize{A,B} &
			\scriptsize{C,D} &  \scriptsize{A,B} & 
			\scriptsize{ } & 
			\scriptsize{A} & \scriptsize{B} &\scriptsize{C,D} &  
			\scriptsize{ } &  \scriptsize{C} &  \scriptsize{D} \\				\hline 
		\end{tabular}
		\begin{tabular}{c}
\\$\,$
		\end{tabular}
		\begin{forest}
			for tree={circle,draw,minimum size=0.5cm,l=1cm,s sep=1cm}
			[\scriptsize{A,B,C,D}
			[\scriptsize{A,B},edge label={node[midway,left,font=\scriptsize]{}}
			[\scriptsize{A,B},edge label={node[midway,left,font=\scriptsize]{}}
			[\scriptsize A,edge label={node[midway,left,font=\scriptsize]{}}]
			[\scriptsize B,edge label={node[midway,left,font=\scriptsize]{}}]
			]			[\scriptsize{},edge label={node[midway,left,font=\scriptsize]{}}
			]			]
			[\scriptsize{C,D},edge label={node[midway,left,font=\scriptsize]{}}
			[\scriptsize{C,D},edge label={node[midway,left,font=\scriptsize]{}}
			[\scriptsize C,edge label={node[midway,left,font=\scriptsize]{}}]
			[\scriptsize D,edge label={node[midway,left,font=\scriptsize]{}}]
			]			[\scriptsize{},edge label={node[midway,left,font=\scriptsize]{}}
			]	]]		;
		\end{forest}
	\tiny
		\caption{QTA worst case with $M=4$}
		\label{fig:wc_qta}
	\end{subfigure}
	\begin{subfigure}[t]{0.49\textwidth}
			\centering
	\begin{tabular}{|c|c|c|c|c|c|}
		\hline
		\scriptsize{A,B,C,D} & \scriptsize{A,B} & \scriptsize{A,B} & 
		\scriptsize{A} & \scriptsize{C,D} &  \scriptsize{C}  \\				\hline
	\end{tabular}
		\begin{tabular}{c}
			\\$\longrightarrow$Time
		\end{tabular}
		\begin{forest}
			for tree={circle,draw,minimum size=0.5cm,l=1cm,s sep=1cm}
			[\scriptsize{A,B,C,D}
			[\scriptsize{A,B},edge label={node[midway,left,font=\scriptsize]{}}
			[\scriptsize{A,B},edge label={node[midway,left,font=\scriptsize]{}}
			[\scriptsize A,edge label={node[midway,left,font=\scriptsize]{}}]
			[\scriptsize B,dotted,edge label={node[midway,left,font=\scriptsize]{}}]
			] [\scriptsize{},dotted,edge label={node[midway,left,font=\scriptsize]{}}
			]			]
			[\scriptsize{C,D},dotted,edge label={node[midway,left,font=\scriptsize]{}}
			[\scriptsize{C,D},edge label={node[midway,left,font=\scriptsize]{}}
			[\scriptsize C,edge label={node[midway,left,font=\scriptsize]{}}]
			[\scriptsize D,dotted,edge label={node[midway,left,font=\scriptsize]{}}]
			]				[\scriptsize{},dotted,edge label={node[midway,left,font=\scriptsize]{}}
			] ]]		;
		\end{forest}
	\tiny
		\caption{SICQTA worst case with $M=4$}
		\label{fig:wc_sicqta}
	\end{subfigure}
	\tiny{\caption{ Worst-case example for Query Tree Algorithms with and without SIC with $M=4$. $u=3$ is set such that maximum number of devices is $N = 2^u = 8$.}}
		\vspace{-0.1cm}
\end{figure*}
\vspace{-0.1cm}
\subsection{Query Tree Algorithm}
\vspace{-0.1cm}

First, we shortly introduce the Contention Tree algorithm. At the start of the algorithm, in binary contention tree algorithm \cite{capetanakis1979tree} any active device sets $c_1 =1$ and transmit. If more than $1$ device is active, the gateway sends a feedback to devices, informing that a collision has happened, and all the active devices do a uniform random selection whether to set $c_2=1$ and $c_3=0$ or vice-versa. The devices that have set $c_2 = 1$ transmit at slot $2$. If again a collision is reported, only those that have transmitted at slot $2$ do a random uniform selection for $c_3$ and $c_4$. Meanwhile, the devices that have previously set $c_3 = 1$, change the values via setting $c_3 =0$ and $c_4=1$. Thus, postponing their transmission. The process goes on until all devices have transmitted successfully.  Even though this process stochastically guarantees that all access codes are unique, the distribution, representing the latency of devices, has a long tail and is not efficient for high reliability constraints.
\begin{algorithm}[!t]
	\caption{Query Tree Algorithm}\label{QTpseudo}
	\begin{algorithmic}[1]
		\Procedure{Generate query}{}
		\State {$Q\gets \lbrace$`0',`1'$\rbrace$}\Comment{Initialize Q list with `0' and `1'}
		\While{$Q$ is not empty}
		\State {$q\gets Q[0]$}\Comment{$q$ is the first element of $Q$}
		\State {Transmit query at the beginning of time-slot}
		\State {Save received packets as $\mathbf{r}$} 
		\State {$f\gets |\mathbf{r}|$}\Comment{Number of received packets}
		\State {$Q.\text{pop}$}\Comment{Delete $Q[0]$}
		\If{$f=0$  or $f=1$}\Comment{Idle or success slot}
		\State{pass}
		\ElsIf{$f>1$}\Comment{Collision slot}
		\State{$Q.\text{append}($`q0',`q1'$)$}
		\EndIf
		\EndWhile
		\EndProcedure{\textbf{End}}
	\end{algorithmic}
	\label{alg:qta}
\end{algorithm}
\par To overcome this issue, Query Tree Algorithm (QTA) is suggested in \cite{choi2007query}. In QTA every device has a unique id formed of $u$ bits. This limits the total number of devices attached to the gateway to $N = 2^u$. In QTA, queries are used instead of feedback but the overhead is the same. In QTA devices are queried with respect to their id bits. The queries start with an empty query. A single bit is appended to the list of queries after each collision, starting from the left-most bit. Each new collision append a new bit. As each device has a unique id, this guarantees that two devices have a unique access decision in worst-case after $u$ transmissions (if all previous $u-1$ bits are the same for two devices). The gateway implementation of QTA is given in Alg.~\ref{alg:qta}, where the device implementation is only answering to the queries matching its id.
\par A detailed example is given for $M=4$ in Fig.~\ref{fig:wc_qta}. We have named the $4$ devices as \{A,B,C,D\} with ids \{000,001,100,101\} respectively. Each circle denotes a slot in the tree. The time-wise progression of the tree is given with slots above the tree. The id size, $u$ is fixed to $3$.

\par In the first slot, 4 devices transmit at the same time and collide. Next slot, the address $0xx$ is queried. Only, A and B transmit. It is again a collision. On the following slot, the query for address $1xx$ is also a collision so the algorithm moves one level down. The address $00x$ is queried and both devices transmit. The query for $01x$ results in an idle slot. Queries for address $001$ and $000$ is done on slot 6 and 7, respectively and both are successes. The algorithm is completed after the process is repeated for right branch.
\vspace{-0.2cm}
\subsection{Query Tree Algorithm with SIC (SICQTA)}
\label{sec:det_tree_sic}
\begin{algorithm}[!t]
	\caption{SICQTA}\label{sicqta_query}
	\begin{algorithmic}[1]
		\Procedure{Generate query}{}
		\State {$Q\gets[\:], q\gets$`0', $k\gets`0$'}\Comment{Initialization}
		\While {$k-1\neq  |Q|$}\Comment{End condition}
		%\State {$q\gets Q[0]$}
		\State {Transmit query at the beginning of time-slot}
		\State {Save received packets as $\mathbf{r}$}
		\State {$q_{\text{b}} \gets [q_1\ldots q_{n-1}\overline{q_n}]$} \Comment{Invert last bit of $q$}
		\State {$f\gets$ $|\mathbf{r}|$} \Comment{Number of received packets}
		%\Comment{Calling function $feedback$}
		\If {$f=0$}\Comment{Idle slot}
		\State {$q\gets$'$q_{\text{b}}$0'} \Comment{Skipping collision}
		\Else
		\State {$Q.\text{append}(q_{\text{b}})$}
		\If{$f>1$}\Comment{Collision slot}
		\State {$q\gets$ `$q0$'}
		\Else
		%\State {$k·\gets $ Number of skipped queries}%{$k\gets kcalculator()$}\\
		%\Comment{Calling function $kcalculator$}
		\Comment{Cancel clean packet and skip.}
		\State {$q\gets Q[-k]+$`0'}
		\State {$Q \gets [Q[0],  \cdots, Q[-k-1]]$} \\\Comment{Skip most recent $k-1$  queries thanks to SIC, $k \geq 1$. }
		%\Comment{$k-1$ queries are omitted due to SIC}
		\EndIf
		\EndIf
		\EndWhile
		\EndProcedure{\textbf{End}}
		%{$*$: $q_{\text{brother}$ is a binary string, it stands for the brother node of $q$ in a ``Query Tree". E.g. if $q=1010$, then the $q_{\text{brother}=1011$. $Q_B$ is a list that contains the $q_{\text{brother}$ from each slot.}\\
	\end{algorithmic}
	\label{alg:sicqta}
	\vspace{-0.08cm}
\end{algorithm}
%QTA can benefit from SIC to improve the throughput and worst-case latency. 

SIC allows recovery of packets from a slot where a collision is observed. If for instance device A and B have transmitted a packet in slot 1, due to collision channel model, the outcome "A+B", is treated as a collision and slot is considered wasted. However, if device B has transmitted its packet in slot 2, the SIC model let us subtract B from "A+B" and enables recovery of A from slot 1.
Instead of breadth first, the SICQTA goes depth-first. After the initial success, it checks if it can cancel the clean packet from previous collisions. If the packet is successfully cancelled then the algorithm skips the direct siblings of those slots.
The algorithmic description of SICQTA is given in Alg.~\ref{alg:sicqta}.\footnote{Open source Python implementation of the algorithms is availabe at: https://github.com/tum-lkn/sicqta}

%SICQTA algorithm consists of 2 procedures: ``generate new query procedure, algorithm \ref{sicqta_query}" and ``Calculate value of k procedure, algorithm \ref{kcal}"

%%K-Calculator
%\begin{algorithm}
%\caption{$k$ calculator}\label{kcal}
%\begin{algorithmic}[1]
%\Procedure{$k=kcalculator$}{}\\
%\Comment{$k$ indicates the omitted slots after applying SIC }
%\State{$k\gets1$, $i\gets 0$}
%\State{$t\gets $ current slot number}
%\Repeat
%\State {$*a\gets received[(t-1)-i]$} 
%\State {$b\gets received[t-i]$}
%\State {$c\gets a-b$}\Comment{Apply Interference Cancellation}
%\If{$c$ is single packet}
%\State {$k\gets k+1$}
%\State {$i\gets i+1$}
%\Else
%\State {\textbf{\text{break}}}
%\EndIf
%\Until{$a$ is the signal from the first slot }
%\State \textbf{return} $k$\\
%\EndProcedure{\textbf{End}} \\
%{$*$: list $received$ contains all the packets received by the gateway. $received[n]$ is the received packets from slot $n$}
%\end{algorithmic}
%\end{algorithm}

%Fuqi: pseudo code end
%This phenomena is adapted for tree algorithms in \cite{yu2005sicta}, where the feedback is expanded from ternary to $k$-ary. However, feedback generation algorithm is not shared in the work and has to be adapted to QTA logic. We also provide the algorithm $k$ generation in this work.
A detailed example for the worst-case behavior of SICQTA is given in Fig.~\ref{fig:wc_sicqta} for $M=4$. In the first slot, all the devices are queried and it is a collision. On the second and third slot, addresses $0xx$ and $00x$ are queried, respectively. Both are collisions. The following slot, $000$ is queried and it is a success. $001$ is not queried, as the gateway recovered the packet from slot 2 and 3. This results in $k=3$ as 2 slots are successfully recovered and this slot is a success. Ids in query list $Q$: $001$ and $01x$ is not queried and skipped. Thus, $10x$ is queried, that results in a collision. Following, $100$ is queried and is a success. The gateway recovered D from slot 5 and the algorithm is terminated.

\vspace{-0.3cm}
\section{Analysis \& Evaluation}
\vspace{-0.1cm}
\label{sec:evalu}
In this section we will evaluate the latency of QTA and SICQTA and give bounds to its performance. We will also compare the performance of our work and \cite{Boyd2018} as we share the same problem definition. Finally, mean delay is compared with state of the art in tree algorithms to show that the stability region is extended. 
\vspace{-0.2cm}
\subsection{QTA}
\vspace{-0.15cm}
\par An upper-bound for latency $y$ of QTA is given in \cite{law2000efficient}:
\vspace{-0.1cm}
\begin{equation}
y \leq { M }\left( u +2 - \text{log} M \right),
\label{eq:qta_bound}
\end{equation}
\vspace{-0.1cm}
where $M$ is the number of active devices. This is a tight bound for $M \ll N$ where with increasing $M$ it has a slack. Using the tree structure we can provide a tighter upper-bound for latency $y$ as,
\vspace{-0.15cm}
\begin{align}
y \leq &  \left\lfloor\frac{M}{2}\right\rfloor 2\left(u +1 - \left\lfloor\text{log}_2 \frac{M}{2} \right\rfloor\right) -1
\label{eq:qta_upbound}
\end{align} 
\vspace{-0.05cm}
Similarly, the tree structure can be used to provide a lower-bound of latency as:
	\vspace{-0.2cm}
\begin{align}
\vspace{-0.1cm}
y &\geq 2M -1.
\label{eq:qta_projected_lower_bound_1}
\end{align} 
	\vspace{-0.08cm}
The proofs are given in App.~\ref{app2} and \ref{app4}, respectively. 

%\\ &\leq  \frac{M}{2} \left( u +4 - \text{log}_2 M \right)-1. 

\par We explain why the example in Fig.~\ref{fig:wc_qta} is the worst-case of a QTA with $M=4$ also shedding light on the proof of the bounds. Four devices are separated into 2 groups of 2 as close as possible to the root of the tree, so they cover as much as non-overlapping slots as possible. Following, devices have repeated the same collision, until the last level of the tree. We observe that for this scenario the total number of slots is $y=11$. Using Eq.~\eqref{eq:qta_bound} we get $13$. This shows that the bound is valid and tight for this setting.

% Reader can observe that if an idle is observed instead of the repeated collisions, as it is guaranteed that the other branch is a collision in that case, that node can be skipped and query can progress to the next level. This is called skipping in tree algorithms and well-known in the state of the art. 

%Contrary to CAC, for QTA the latency given with Eq.~\eqref{eq:qta_bound} is an upper bound that occurs only with $M$ devices that have an address with a single bit difference. In Fig.~\ref{fig:bounds} we have plotted the number of devices $N$ (x-axis) versus the maximum achievable latency constraint $\text{L}$ (y-axis) given $M = \{2,4,8\}$ active devices. An interesting first observation is that CAC and QTA with $M=2$ behaves almost the same as QTA, even though QTA uses extra feedback. We have an intuitive projection from this result, which is that feedback does not help for the worst case. 

\subsection{SICQTA}
Intuitively, the efficiency of the \cite{yu2005sicta} comes from the possibility to skip some slots in the tree. As it is shown in \cite{yu2005sicta}, the throughput of BTA is doubled. However, the throughput is the expected number of slots and this result cannot be directly translated to worst-case latency of SICQTA from QTA. We have to adapt the Eq.~\eqref{eq:qta_bound} for SICQTA using the skipping capability of SIC. 
%Initially, we include the skipping of the idle slots in the worst-case, 
%\begin{align}
%y &\leq  \left\lfloor \frac{M}{2} \right\rfloor \left( u +4 -\left\lfloor \text{log}_2 M \right\rfloor\right)-1 .
%\label{eq:qta_projected_bound_13}
%\end{align}
The total number of skipped slots $\mathbf{S}$ compared to worst-case of QTA, given $M$ active devices can be written as,
\vspace{-0.1cm}
\begin{equation}
\mathbf{S}  = \left\lfloor\frac{M}{2} \right\rfloor \left( u-1 - \left\lfloor\text{log}_2 \frac{M}{2} \right\rfloor \right) +\sum_{i=1}^{\lfloor\text{log}_2 M\rfloor} \left\lfloor\frac{M}{2^i}\right\rfloor.
\label{eq:shrink}
\end{equation} 
The proof is given in \ref{app3}.
%\mg{Think about a more formal proof.}

We can use this finding to provide an upper-bound for latency of SICQTA using Eq.~\eqref{eq:qta_upbound} and removing the skipped slots,
\vspace{-0.1cm}
\begin{equation}
y \leq \left\lfloor \frac{M}{2} \right\rfloor \left( u +4 -\left\lfloor \text{log}_2 M \right\rfloor\right)-1- \sum_{i=1}^{\lfloor\text{log}_2 M\rfloor} \left\lfloor\frac{M}{2^i}\right\rfloor.
\label{eq:sicqta_bound}
\end{equation}
Intuitively, the algorithm needs at least $M$ slots for $M$ active devices and a lower-bound for latency of SICQTA can be given as $
y \geq M.$
\par This is given without any proof, as in best-case no repetition occurs such that every slot is recoverable from another.
\par The upper-bound for latency can be used for the throughput calculation of the SICQTA. If number of active devices is the same as the number of total devices, i.e., $M = N = 2^u$. Then we expect SICQTA to have a throughput of 1, as each slot in the tree should be different from one another.

% we use the tree structure to calculate the number of leaves in the full tree for QTA. We know that given all devices are active, all leaves of the QTA has to be traversed. The total number of slots is equal to the total number of leaves in a binary tree with $u$ levels. 
%	\begin{figure}[t!]
%		\includegraphics[width=0.5\textwidth]{images/compare_number_of_users2}
%		\caption{QTA with $M = \{2,4,8\}$ compared against CAC with $M=2$. The upper bound of latency can be calculated given the number of devices $N$. The reliability constraint is set to $R=1$. }
%		\label{fig:bounds}
%	\end{figure}

Eq.~\eqref{eq:sicqta_bound} is a relaxed bound, but it becomes tight for integer values of $\text{log}_2 M$. Plugging in $M = 2^u$ we get,
\begin{align}
2^u \leq y \leq 2^{u+1} -1 - 2^{u+1} + 2^{u} + 1 = 2^u.
\end{align}
Thus, we have a throughput of $1$ as expected. The proof is given in App.~\ref{app1}.
% For level $m$ we have $2^m$ leaves in the tree summing this up for all levels up to $u$, the total number of leaves in a binary tree with $u$ levels is
%\begin{equation}
%	\mathbf{T} = \sum_{m=0}^u 2^m = 2^{u+1} -1.
%	\label{eq:total_nodes}
%\end{equation}
%For Eq.~\eqref{eq:shrink} we replace $M$ with $2^u$ and we get,
%
%\begin{equation}
%\mathbf{S}  = 2^{u+1} - 2^{u} -1.
%\label{eq:shrink_tot}
%\end{equation}
%The proof is given in App.~\ref{app1}. For the number of slots in SICQTA with $M = 2^u$ we have $y =  \mathbf{S}  - \mathbf{T} $. Plugging the calculated results we get $y = 2^u$. 

We can check the bound via the example in Fig.~\ref{fig:wc_sicqta}.  We see that in total $6$ slots are used for SICQTA in the example. Using Eq.~\eqref{eq:sicqta_bound} we get $6$ showing that the bound is valid and tight for this scenario.
\par In Tab.~\ref{tab:sic} we have compared the number of devices $N$ supported by CAC-SIC with SICQTA. The number of active devices are fixed to $M=3$ for CAC, because these are the only available results in \cite{Boyd2018}. For SICQTA, we see that with relaxed delay constraint the number of devices supported increases exponentially. And even though the results are similar for low latency constraints, the difference increases with increasing $\text{L}$. Also the results for SICQTA is easily extensible to other $M$ values, while an exhaustive search is required to build codes for CAC-SIC. On the other hand effect of feedback is neglected in this analysis.
	%\begin{figure}[t!]
	%	\includegraphics[width=0.5\textwidth]{images/sic_qta_latency}
	%	\caption{Latency bounds for QTA and SICQTA. The new bounds are compared to previously given bound for QTA. Excessive simulations show the validity of the bounds. The maximum number of levels is set to $u=6$ and $M$ is varied (x-axis). }
	%	\label{fig:bounds_compare}
	%\end{figure}
	
\subsection{Simulations}
We have done Monte Carlo experiments on a python based discrete event simulator $10^6$ samples for each experiment varying the number of active devices.

\begin{figure}[t!]
	\vspace{0.1cm}
	\begin{subfigure}{0.48\textwidth}
		\includegraphics[width=1\textwidth]{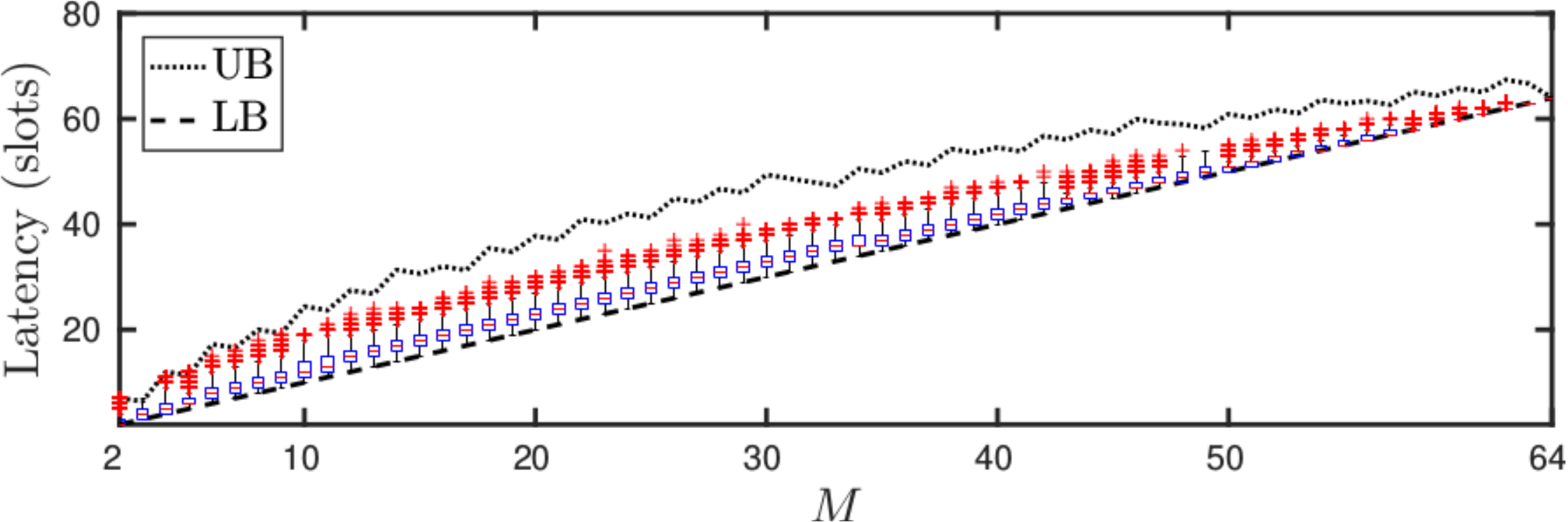}
		\caption{Latency}
		\label{fig:bounds_comparesicqta}
	\end{subfigure}
	\begin{subfigure}{0.48\textwidth}
		\includegraphics[width=1\textwidth]{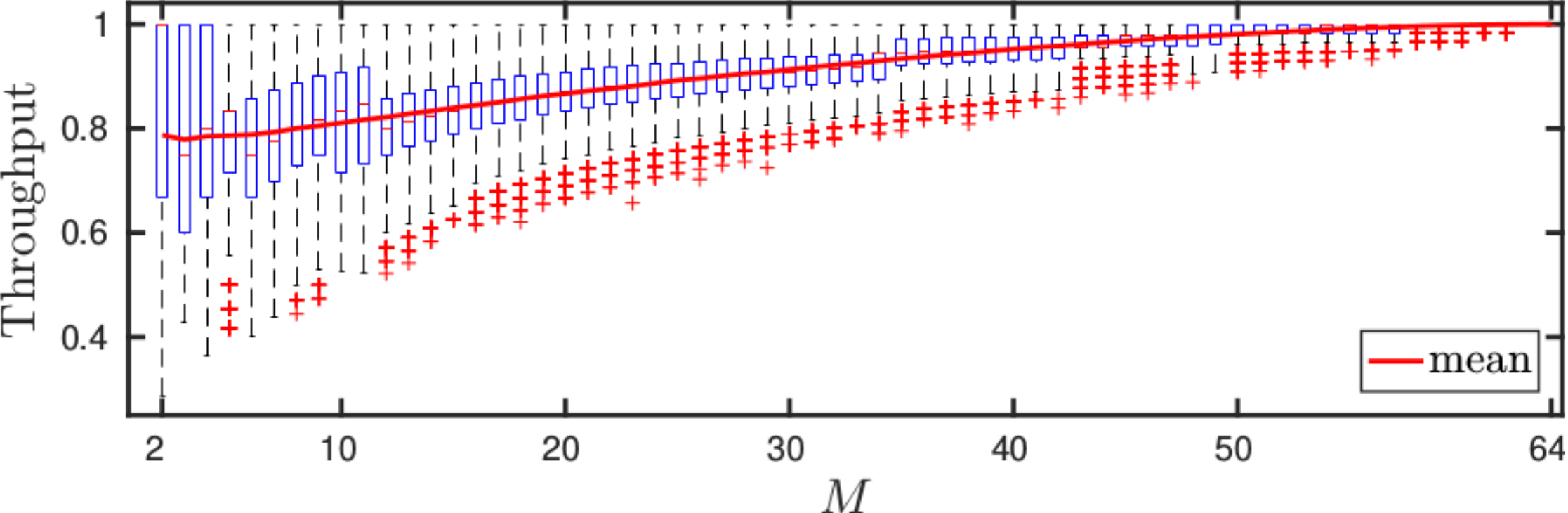}
		\caption{Throughput}
		\label{fig:bounds_tpt_comparesicqta}
	\end{subfigure}
	\small \caption{Excessive simulations show the validity of the bounds. The maximum number of levels is set to $u=6$, $N=64$ and $M$ is varied (x-axis). }
	\label{fig:bounds_compare_sicqta}
	\vspace{-0.3cm}
\end{figure}

In Fig.~\ref{fig:bounds_compare_sicqta} we have plotted the bounds versus simulation for SICQTA. x-axis depicts the varying active number of devices $M$ and the y-axis presents the latency. %We see that the new bound provides a tighter bound especially for increasing $M$.
We have set $u=6$ so implicitly $N=64$, and we have varied the number of active devices $M$. We see that with $10^4$ iterations for each data point in simulations the bounds are never surpassed and the difference between the lower and the upper bound is quite low.

\par As we deal with worst-case latency, this is the latency of the last device. In Fig.~\ref{fig:bounds_tpt_comparesicqta} we have evaluated the throughput with varying active number of devices $M$. Mean throughput is almost always above $0.8$ while the tail is also quite constrained, especially with increasing $M$. 	
%	\mg{to be extended with further simulation results with increasing $u$.}	
%	\mg{some simulations or some result with finite length IRSA or frameless should be found.}

	In Fig.~\ref{fig:delay_comp_algs} we extend the delay vs throughput comparison in \cite{yu2005sicta} with SICQTA. In this simulation scenario continous arrivals are considered. If a device gets a packet to transmit while there is an on-going resolution, the device is queued until the end of that resolution, reflecting the setting in \cite{yu2005sicta}. We see that SICQTA enables a new throughput region that extends to throughput of $0.93$ with $u = 4$. Also with $u=6$ the throughput with stable latency is around $0.86$. Of course SICQTA becomes similar to SICTA with increasing $u$ value. This is logical as SICTA can be considered as a special setting of SICQTA with $u = \infty$. Here, it is shown that with $u=10$ the behavior is almost the same as SICTA. It is worth mentioning that the average resolution time is increased as we see a shift on the y-axis compared to SICTA. We have also simulated higher values of $u$, i.e., $u=16$ and did not observe any difference so they are not plotted here to avoid clutter. For decreasing $u$ the throughput is expected to increase further reaching $1$.

\vspace{-0.2cm}
\section{Discussions}
\label{sec:disc}
%\par Moreover, we see that both algorithms are also suitable for different scenarios such as Massive IoT as the $N$ increases exponentially after a certain latency constraint. On the other hand, lower latency constraints is feasible given $\frac{M}{N}$ ratio is low. This makes CAC a good candidate for cyber-physical systems as long as $M$ can be constrained. One important advantage of QTA is that its access code design does not depend on $M$.% and feedback speeds up the algorithm.
\begin{table}[!t]
	\centering
	\begin{tabular}{|c|c|c|c|c|}
		\hline
		Constraint & L$=4$ & L$=5$ & L$=6$ & L$=7$\\
		\hline
		CAC-SIC \cite{Boyd2018} & $7 $& $11$ &$ -$ &$ -$ \\				\hline 
		SICQTA $M=3$& $8$ &  $16$ & $32 $& $64$\\ \hline
		SICQTA $M=4$& $4$ &  $8$ & $8 $& $16$\\				\hline 
	\end{tabular}
	\caption{Number of devices supported by CAC-SIC for fixed number of active devices $M=3$, with varying latency constraint, compared to SICQTA.}
	\label{tab:sic}
	\vspace{-0.4cm}
\end{table}
One important point for SICQTA compared to QTA is that the knowledge of $M$ does not improve the upper-bound of latency. The knowledge of $M$ would be used in this case to skip to level $\lfloor\text{log}_2 M\rfloor$. However, in the worst-case all collisions happening before this level consist of different devices, and under a SIC framework, they can all be recovered from each other to obtain useful slots. So the number of skipped slots with knowledge of $M$ would be equal to those skipped due to SIC. However, application of knowledge of $M$ to QTA can improve the worst-case performance and bring it close to SICQTA.

\par We have compared the feedback based algorithms to non-feedback based algorithms here. However, we assumed that the feedback is instantaneous and costless. In reality that is not the case. The latency incurred due to transmission and reception may even involve hardware delays such as switching from transmit to receive and vice-versa. We leave this open for future work. 

%\par Another interesting line of work could be simulating QTA and saving the code $\mathbb{c}$ generated by each device in the worst-case. Use of the codes generated by the worst-case can be a solution to design access codes without feedback. This is related to our earlier projection that with a fixed $M$ the feedback does not help for the worst-case. 

\par We are also working on prototyping this algorithm through IEEE 802.15.4 capable sensors and SDRs. One observation we have is that depending on the quality of the sensor device, the phase noise accumulates through successive interference cancellation and this makes collisions of 6 packets, a wasted slot as cancellation fails due to accumulated phase noise. Algorithmic solutions, such as starting the queries from level $u-3$\footnote{This will cap the maximum number of collided devices to $8$.}, should be considered to overcome such hardware constraints. Curious reader can refer to \cite{4276942}  for a theoretical model that incorporates variances in the hardware to the SIC capacity and to \cite{zhou2006models} for practical characterization of causes for hardware variances.

	\vspace{-0.2cm}
\section{Conclusion}
\label{sec:conc}

In this work we have evaluated the problem of uplink resource allocation for unknown number of active devices. We believe that this problem represents the important uplink resource allocation problem for multiple control loops sharing the same wireless network. As a solution we present the algorithm Successive Interference Cancellation Query Tree Algorithm (SICQTA). The advantage of the algorithm compared to previous algorithms is the high-throughput performance and the hard latency guarantees. The bounds for the performance are proven analytically and further validated with simulations. 

\par Future work can investigate relaxing the assumptions made for easy investigation of the protocol. Firstly, the feedback is assumed instantaneous and costless, accumulation of feedback messages should be considered to decrease this bottleneck as much as possible. Secondly, we assumed that SIC works perfectly. However, due to accumulated phase noise some collisions cannot be recovered via SIC and indeed result in wasted slots. This should be evaluated and incorporated into the protocol design. Thirdly, even though it is intuitive that decreasing latency and increasing reliability helps for the cyber-physical systems, an integrated evaluation of control and communication should be done to provide concrete results.

 \vspace{-0.1cm}
\appendix
 \vspace{-0.1cm}
\subsection{Proof for upper-bound for latency of QTA}
\label{app2}
\begin{figure}[t!]
 	\centering
 	\includegraphics[width=0.38\textwidth ]{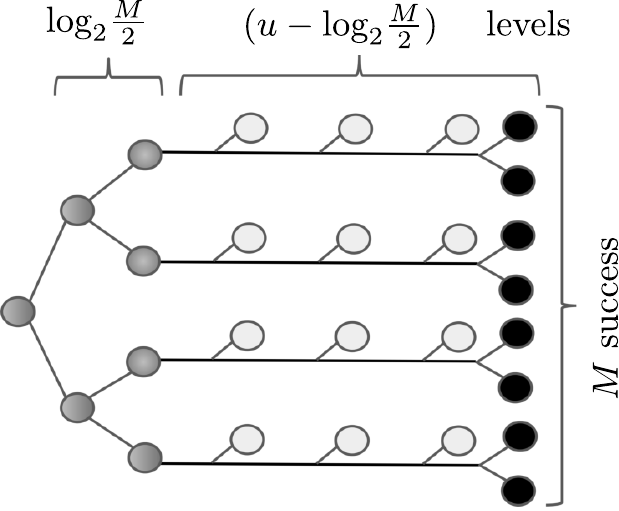}
 	\small
 	\caption{The worst-case tree structure for Query Tree Algorithm.}
 	\label{fig:explain}
 	 \vspace{-0.4cm}
 \end{figure}

The worst-case for QTA is illustrated in Fig.~\ref{fig:explain}. 
An intuitive explanation is as follows: A device can re-transmit at maximum $u$ times in the worst-case as that is the size of addresses and every device has a unique address. In this case the device is successful with the $u^\text{th}$ transmission and it has experienced $u-1$ collisions. In order to have a collision we need at least $2$ devices, and at the worst-case all devices are grouped into two, thus $\lfloor\frac{M}{2}\rfloor$ groups. Each group collides separately for $u-1$ times, where there will be idles on the unexplored slots so $2\cdot(u-1)$ slots, followed with 2 transmissions for success of each device, we get 
\begin{equation}
y \leq \left\lfloor\frac{M}{2}\right\rfloor 2\cdot(u-1 +1)
\vspace{-0.2cm}
\end{equation}
 slot uses in total. We take into account, the activity of the groups of two only after the level $\left\lfloor\text{log}_2 \frac{M}{2}\right\rfloor$. As the initial levels have a lot of overlap, we can remove these levels and consider them separately as
\begin{equation}
y \leq  \left\lfloor\frac{M}{2}\right\rfloor 2\cdot\left(u - \left\lfloor\text{log}_2 \frac{M}{2} \right\rfloor\right)  + \mathbf{R},
\label{eq:on_proof}
\end{equation}  where $\mathbf{R}$ represents the overlapping slots. The number of overlapping slots can be calculating by summing the total number of slots up to level $\text{log}_2 \frac{M}{2} \geq \left\lfloor\text{log}_2 \frac{M}{2} \right\rfloor $ of the tree. We can calculate the total number of nodes in this upper part of the tree as,
\begin{equation}
\mathbf{R} \leq  2^{m+1} -1= 2^{\text{log}_2 \frac{M}{2} +1} -1 = M -1 .
\end{equation}
Plugging this in Eq.~\eqref{eq:on_proof} we get,
\begin{align}
y \leq & \left\lfloor\frac{M}{2}\right\rfloor 2\cdot\left(u - \left\lfloor\text{log}_2 \frac{M}{2} \right\rfloor\right)  + M -1, \\ \leq &  \left\lfloor\frac{M}{2}\right\rfloor 2\cdot\left(u +1 - \left\lfloor\text{log}_2 \frac{M}{2} \right\rfloor\right) -1.
\label{eq:on_proof2}
\end{align} 
\begin{figure}[t!]
		\vspace{0.1cm}
	\centering
	\includegraphics[width=0.44\textwidth ]{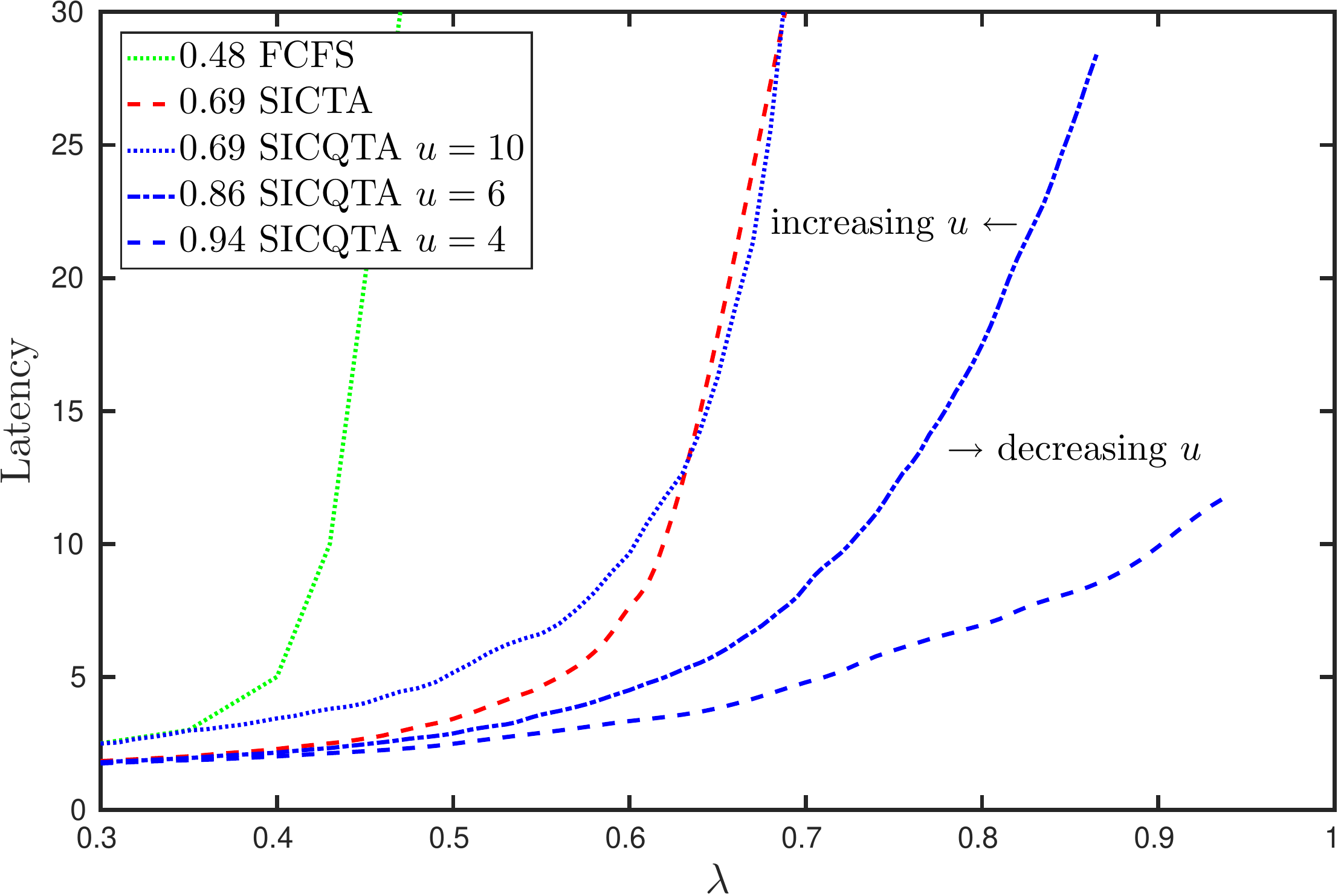}
	\small \caption{Delay vs throughput of feedback based random access algorithms.}
	\label{fig:delay_comp_algs}
		\vspace{-0.1cm}
\end{figure}
	\vspace{-0.4cm} 
\subsection{Proof for lower bound for latency of QTA}
\label{app4}
The best-case in the tree with $M$ devices, is that they are organized as a triangle, guaranteeing they are as close as possible to the root. So the level of the successes are almost the same. However, the level of the devices can be the same only if $\text{log}_2 M$ is an integer. If it is not an integer, the best-case would be some of the devices are successful at level $l_1 = \lceil \text{log}_2 M \rceil$ and the others are at $l_2 = \lfloor \text{log}_2 M \rfloor$. In order to have a complete triangle we would need that devices at level $l_2$ would each have 2 children at $l_1$. So the number of slots at $l_1$ is equal to the sum of number of devices at $l_1$ plus twice the number of devices at $l_2$. The number of slots at a level can also be written as $2^l$ so we can write,
\begin{align}
2^{l_1}   = M_{l_1} + 2\cdot M_{l_2}
\label{eq:lb_proof1}
 \vspace{-0.2cm}
\end{align}
where $M_{l_1}$ and $M_{l_2}$ is the number of devices successful in $l_1$ and $l_2$ respectively. We know that the total number of devices is $M = M_{l_1} +M_{l_2}$. so we can re-write Eq.~\eqref{eq:lb_proof1} as
\begin{align}
M_{l_1} = 2\cdot M - 2^{l_1}.
\label{eq:lb_proof2}
 \vspace{-0.2cm}
\end{align}

If we do not consider the level $l_1$, the tree is a full triangle up to level $l_2$. We can calculate the total number of slots in the tree for the best-case $y_{LB}$ through calculating the number of slots for the full tree up to $l_2$ and adding $M_{l_1}$
\begin{align}
y_{LB}  = 2^{l_2 +1} -1 + M_{l_1}.
 \vspace{-0.2cm}
\end{align}
By definition of flooring and ceiling operation $l_2 +1 = l_1$ if $\text{log}_2 M$ is not an integer. And we can plug Eq.~\eqref{eq:lb_proof2} in to get,
\begin{align}
y_{LB}  = 2^{l_1} -1 +2\cdot M - 2^{l_1} = 2\cdot M -1.
 \vspace{-0.2cm}
\end{align}
 When $\text{log}_2 M$ is an integer the lower-bound is directly given with $2^{\text{log}_2 M +1}-1$, which is equal to the result so we do not mention it separately. 
 
\vspace{-0.1cm}
\subsection{Proof for number of skipped slots $\mathbf{S}$}
\label{app3}
The skipping in SICQTA consists of two different parts. First part is skipping the idles $S_I$ and second part is skipping the canceled slots $S_C$. 
So we can write $\mathbf{S}  = S_I +S_C$.

The upper-bound for latency of QTA is derived using groups of 2 devices sticking together until the last level of the tree. At the last level they transmit separately, each as a success. The idles occur after separation from the top triangle until the end of the tree. We have $\left\lfloor\frac{M}{2} \right\rfloor$ collisions and the number of levels until the end of the tree gives us the number of skipped idle slots as 
 \vspace{-0.15cm}
\begin{equation}
	S_I = \left\lfloor\frac{M}{2} \right\rfloor \left( u-1 - \left\lfloor\text{log}_2 \frac{M}{2} \right\rfloor \right) .
\end{equation}
 \vspace{-0.15cm}

Thanks to SIC, after one success the other device does not have to transmit anymore, as after one success the other device can be recovered from the previous collision. Thus, at least $\frac{M}{2}$ slots are skipped for the last level of the tree. 

\par This skipping can be applied to also formation of groups of 2. Groups of 2 are formed from groups of 4. Thus, for the first group formed out of 4 devices, the other group can be recovered from the collision, so one slot can be saved for each separation. In this step we can save $\frac{M}{4}$ slots. This logic can be extended up to $\lceil \text{log}_2 M \rceil $ separations as we have a binary splitting process. This gives us,
 \vspace{-0.2cm}
\begin{equation}
S_C   = \sum_{i=1}^{\lfloor\text{log}_2 M\rfloor} \left\lfloor\frac{M}{2^i}\right\rfloor.
\label{eq:shrinked}
\vspace{-0.2cm}
\end{equation}
\vspace{-0.15cm}
Finally, we can write,
\vspace{-0.15cm}

\begin{equation}
\mathbf{S}  = \left\lfloor\frac{M}{2} \right\rfloor \left( u-1 - \left\lfloor\text{log}_2 \frac{M}{2} \right\rfloor \right) +\sum_{i=1}^{\lfloor\text{log}_2 M\rfloor} \left\lfloor\frac{M}{2^i}\right\rfloor.
\vspace{-0.05cm}
\end{equation} 
 \vspace{-0.15cm}

\subsection{Proof for number of skipped slots with $M=2^u$}
\label{app1}
We plug in $M= 2^u$ to Eq.~\eqref{eq:shrinked}
 \vspace{-0.15cm}
\begin{equation}
\mathbf{S}  =  \left\lfloor\frac{2^u}{2} \right\rfloor \left( u-1 - \left\lfloor\text{log}_2 \frac{2^u}{2} \right\rfloor \right) +  \sum_{i=1}^{\lfloor\text{log}_2 2^u\rfloor} \left\lfloor\frac{2^u}{2^i}\right\rfloor = \sum_{i=1}^{ u} {2^{u-i}}.
\end{equation}
 \vspace{-0.15cm}

\begin{multline}
\mathbf{S}   = \sum_{i=1}^{u} {2^{u-i}} = 2^u \left(\sum_{i=0}^{u-1} 2^{-i} -1 +2^{-u} \right) \\
  = 2^u \left(\frac{1-2^{-u}}{1-2^{-1}} -1 +2^{-u} \right) 
  = 2^{u+1} -2^{u}-1,
  \label{eq:ter}
\end{multline}
\vspace{-0.15cm}
is what we get, as $u$ is the number of maximum levels and is an integer we can remove the floor operation. So we can plug Eq.~\eqref{eq:ter} in Eq.~\eqref{eq:sicqta_bound} to get,
\begin{equation}
y \leq \left\lfloor \frac{2^u}{2} \right\rfloor \left( u +4 -\left\lfloor \text{log}_2 2^u \right\rfloor\right)-1- 2^{u+1} +2^{u}+1.
\label{eq:sicqta_bound_proof}
\end{equation}
\vspace{-0.4cm}

\bibliographystyle{IEEEtran}
\bibliography{query_tree}

% that's all folks
\end{document}